\documentclass[english]{aastex}
\usepackage{amssymb}
\usepackage{graphicx}
\usepackage{babel}
\begin{document}

\title{On the existence of regular and irregular outer moons orbiting the Pluto-Charon system }

\author{Erez Michaely, Hagai B. Perets and Evgeni Grishin}
\affil{Physics Department, Technion - Israel Institute of Technology, Haifa
3200004, Israel}

\begin{abstract}
The dwarf planet Pluto is known to host an extended system of five
co-planar satellites. Previous studies have explored the formation
and evolution of the system in isolation, neglecting perturbative
effects by the Sun. Here we show that secular evolution due to the
Sun can strongly affect the evolution of outer satellites and rings
in the system, if such exist. Although precession due to extended
gravitational potential from the inner Pluto-Charon binary quench
such secular evolution up to $a_{crit}\sim0.0035$ AU ($\sim0.09$
$R_{Hill}$ the Hill radius; including all of the currently known
satellites), outer orbits can be significantly altered. In particular,
we find that \emph{co-planar} rings and satellites should not exist
beyond $a_{crit}$; rather, satellites and dust particles in these
regions secularly evolve on timescales ranging between $10^{4}-10^{6}$
yrs, and quasi-periodically change their inclinations and eccentricities
through secular evolution (Lidov-Kozai oscillations). Such oscillations
can lead to high inclinations and eccentricities, constraining the
range where such satellites (and dust particles) can exist without
crossing the orbits of the inner satellites, or crossing the outer
Hill stability range. Outer satellites, if such exist are therefore
likely to be \emph{irregular} satellites, with orbits limited to be
non-circular and/or highly inclined. These could be potentially detected
and probed by the New-Horizon mission, possibly providing direct evidence
for the secular evolution of the Pluto satellite system, and shedding
new light on its origins. 
\end{abstract}

\section{Introduction}

Since its discovery in 1930 \citep{tom46} the dwarf planet Pluto
has provided us with valuable information on the origin of the Solar
system, its structure, its dynamical evolution and its basic building
blocks. It was the first object to be discovered in the Kuiper belt
and suggest its existence; its orbital resonance with Neptune and
high inclination provided strong evidence for Neptune migration \citep{mal93}
, and thereby provided important clues into the early history of the
Solar System and its assemblage. The discovery of its close companion
Charon \citep{chr+78} revealed the existence of a new type of planetary
systems, binary planetesimals/dwarf-planets, and its close and tidally-locked
configuration provided further clues for its collisional origin \citep{can05}. 

A series of discoveries in recent years showed the existence of no-less
than four additional satellites, including Styx, Nix, Kerberos and
Hydra orbiting Pluto on circular orbits with semi-major axes ranging
between $\sim$42K-65K km, i.e. residing between $\sim0.007-0.011$
of the system Hill radius, $R_{Hill}=a_{P}(1-e_{P})(m_{P}/3m_{\odot})^{1/3}$;
where $a_{P}$, $e_{P}$ and $m_{P}$ are the semi-major axis (SMA)
of the orbit of Pluto around the Sun; the orbital eccentricity; and
the mass of Pluto, respectively; $m_{\odot}$ is the mass of the Sun
(\citealp{bro+15} and references therein). These recent findings
gave rise to a series of studies on the formation, stability and evolution
of this extended system \citep{bro+02,lee+06,ste+06,war+06,lit+08,lit+08b,can11,you+12,giu+13,che+14b,gio+14,ken+14,bro+15b,giu+15,por+15},
that could shed light not only on the origin of the Pluto system,
but also provide new clues to the understanding of the growth of planetary
system in general, as well as on migration and resonant capture processes
in planetesimal disks. Better understanding of the properties of the
Pluto system and its constituents is therefore invaluable for advancing
our knowledge of the Solar System, its building blocks and and its
origins. The New-Horizon mission, launched in 2006, was designed to
explore the Pluto-Charon system and help accomplish these goals \citep{you+08}.
New horizon is now reaching its climax, as it makes its close flyby
near the Pluto-Charon system, aiming to characterize Pluto and its
satellites at an unprecedented level, and potentially detect additional
satellites and/or planetesimal rings in this system. Understanding
the dynamical history and the current configuration of the Pluto satellite
system system are therefore invaluable for realizing New-Horizons'
data collecting and characterization potential, and use them in the
interpretation of the data and their implications for the origins
of Pluto and the Solar System history. 

Here we show that secular evolutionary processes due to gravitational
perturbations by the Sun, previously neglected, may have played a
major role in the initial formation of the Pluto system as well as
in the later evolution of its satellite system and its configuration
to this day. As we discuss below, our finding provide direct predictions
for the possible orbits of outer moons in the Pluto-Charon system,
if such exist. In particular we show that co-planar planetesimal rings
can not exist beyond a specific critical separation from Pluto-Charon
(significantly smaller than the Hill radius of the system), and that
any moon residing beyond this critical separation must be an irregular,
inclined and/or eccentric moon. Moreover, we constrain the orbital
phase space regimes (semi-major axis, eccentricity, inclination) in
which dust, planetesimal-rings and/or outer moons can exist on dynamically
and secularly stable orbits.

\section{Secular evolution of sattelites in the Pluto-Charon system}

Charon is Pluto's closest and largest moon, with semi-major axis of
$a_{pc}\sim19.4K$ km (period of $6.387$ days) and mass $\sim1.52\times10^{24}$g
(mass ratio of $\mu=m_{c}/m_{P}=0.116$). It revolves around Pluto
on a tidally-locked (i.e. in double synchronous state) circular orbit
with an inclination of $I\sim119^{\circ}$ in respect to the orbit
of Pluto around the Sun (which in itself is inclined in an angle of
$\sim17^{\circ}$ in respect to the ecliptic plane). Together with
the Sun, the Pluto-Charon system comprises a hierarchical triple system,
with Pluto and Charon orbiting each other in an inner binary orbit,
and their center of mass orbiting the Sun in an outer binary orbit.
The evolution of such hierarchical gravitating triple systems may
give rise to secular dynamical processes, in which the outer third
body perturbs the orbit of the inner binary, leading to secular precession
of the inner binary, typically exciting its eccentrcity. In particular,
triples with high mutual inclinations between the inner and outer
orbits (typically $>\sim40^{\circ}$, and somewhat lower for eccentric
systems) are subject to quasi-periodic secular evolution, so called
Lidov-Kozai (LK) cycles \citep{lid62,koz62}, which could affect Pluto's
satellites as well as other Solar System binary planetesimals \citep{per+09}
and satellites of the gas-giants \citep{car+02}. Such LK evolution
leads to the inner orbit precession and in turn gives rise to high
amplitude eccentricity and inclination oscillations that occur on
the LK precession timescales, which are much longer than the orbital
period.

In principle, the high mutual inclination of the Charon-Pluto system
makes it highly susceptible to such LK secular evolution, and in the
absence of other forces it would have lead to a collision between
Pluto and Charon. However, such processes are highy sensitive to any
additional perturbations of the system precession. In particular,
the effects of the tidal forces between Pluto and Charon at their
current separation lead to a significant precession of their orbital
periapse, which will be denoted $\dot{g}_{PC}$. This, in turn, quenches
any LK secular evolution, and keeps Pluto and Charon on a stable circular
orbit, with no variations of the orbital eccentricity and the mutual
inclination in respect to the orbit around the Sun. 

The strong dependence of the tidal forces on the satellite separation
renders the tidal effects on Pluto's four other known satellites negligible.
However, periapse precession can still be induced by the non -- point-mass
gravitational-potential of the inner Pluto-Charon binary \citep{lee+06,ham+15},
similar to the case of circumbinary planets \citep{ham+15b,mar+15,mun+15}.
In this case, if the inner-binary induced precession timescale, $\tau_{PC}$,
is shorter than the LK precession timescale, $\tau_{LK}$, the orbit
will not librate and the LK secular evolution will be quenched. If,
on the other hand, LK period is shorter than the precession timescale,
the binary-induced precession is slow and the orbit has sufficient
time to librate and build-up significant LK oscillations. The critical
SMA at which these timescales become comparable is given by (Fig.
1; see appendix for full derivation)
\begin{equation}
a_{crit}=\left[\frac{3}{8}\frac{a_{PC}^{2}a_{P}^{3}m_{P}m_{C}\left(1-e_{p}^{2}\right)^{3/2}\left(5\theta^{2}-1\right)}{\left(m_{P}+m_{C}\right)\left(1-e_{m}^{2}\right)^{2}m_{\odot}}\right]^{1/5},\label{eq:a_crit-1}
\end{equation}
where parameters with sub-indexes PC correspond to the orbital parameters
of the Pluto-Charon orbit, and the $m$ index refers to the orbital
parameters of the orbiting moon. The inner orbit semi-major axis (SMA)
for Pluto-Charon mutual orbit is $a_{PC}$; the SMA of the moon orbit
around Pluto-Charon is given by $a_{m}$; the eccentricity of the
inner orbit, is $e_{PC};$ the eccentricity of the outer orbit is
$e_{m}$; the arguments of the pericenter of the inner and outer orbits
are $g_{PC},\, g_{m}$, respectively and $\theta\equiv\cos i$ is
the cosine of the inclination between the two orbit planes denoted
by $i$ . Charon mass is given by $m_{C}$ and the mass of the Sun
is denoted by $m_{\odot}$. 

\begin{figure}
\includegraphics[scale=0.5]{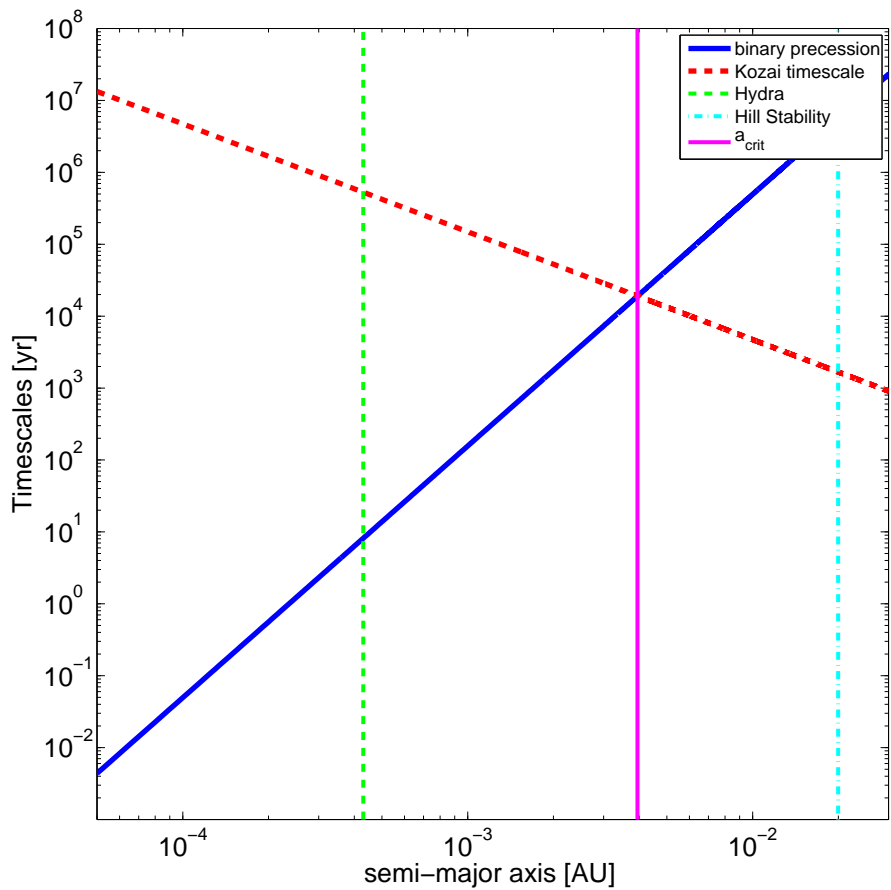}

\caption{{\footnotesize{\label{fig:timescales}Comparison between the precession
timescales for a satellite orbit induced by perturbations by the Sun
(Lidov-Kozai evolution) and the precession induced by the inner Pluto-Chaon
binary. Beyond the critical separation,$a_{crit}$, at which the timescales
become comparable, satellites/rings become highly susceptible to Lidov-Kozai
secular evolution. }}}
\end{figure}

\begin{figure}
\includegraphics[scale=0.25]{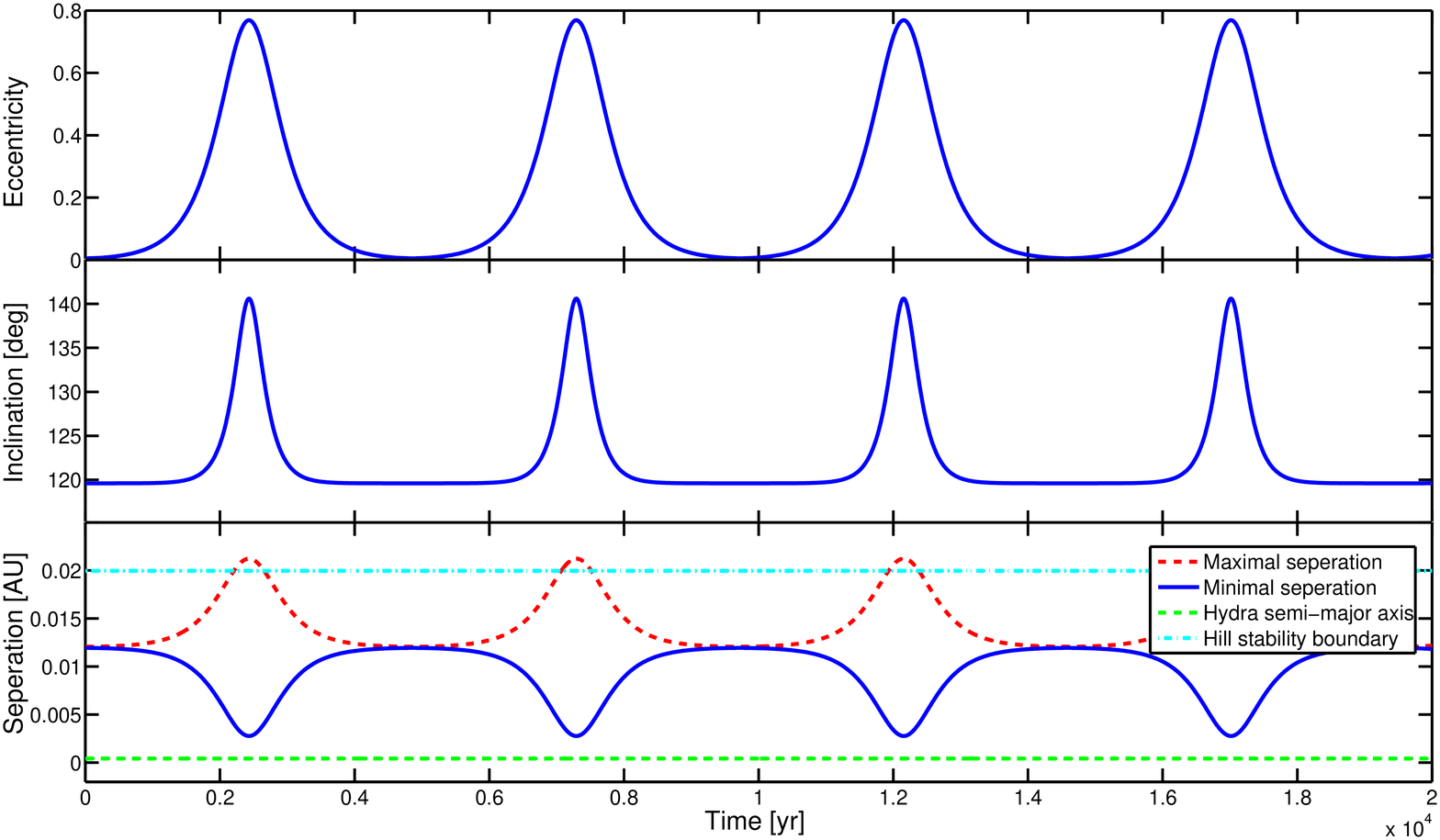}

\includegraphics[scale=0.25]{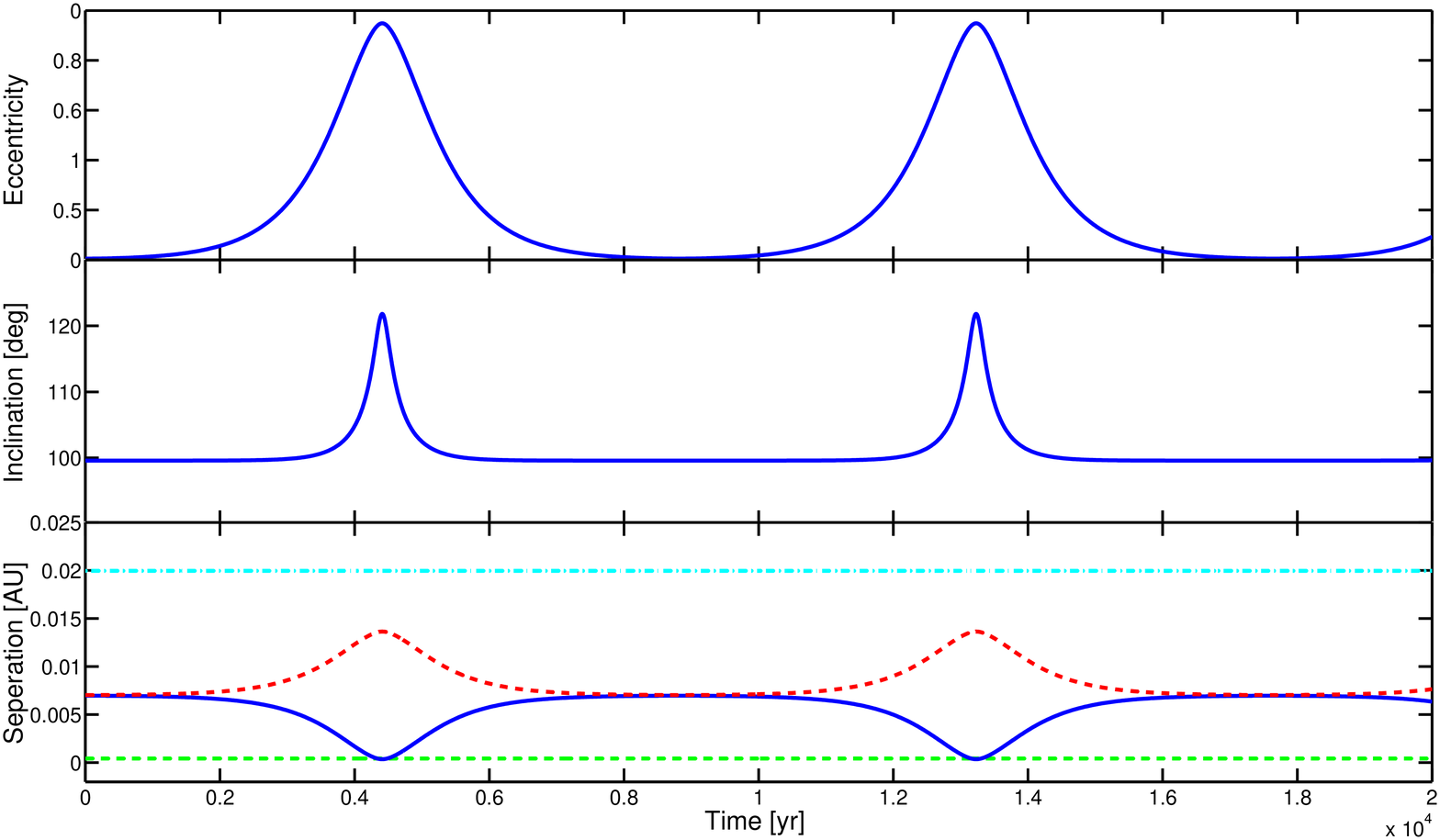}\caption{{\footnotesize{\label{fig:kozai-examples}Examples for the secular
evolution of outer satellites in the excluded regions. Top: The evolution
of a satellite on an initially co-planar (with respect to the Pluto-Charon
orbit;$119^{\circ}$ with respect to the orbit of Pluto around the
Sun), circular orbit. The eccentricity and inclination of the orbits
change periodically due to LK evolution until the satellite separation
becomes too large and crosses the Hill stability region, making the
satellite orbit unstable. Right: The evolution of a satellite on an
initially inclined (-20 degrees in respect to Pluto-Charon; $99^{\circ}$
with respect to the orbit of Pluto around the Sun), circular orbit.
The eccentricity and inclination of the orbits change periodically
due to LK evolution until the satellite separation becomes too small
and crosses the orbit of Hydra; at which point the satellite system
is likely to destabilize and/or the satellite can collide with Hydra;
such satellite are therefore unlikely to exist in the Pluto-Charon
system.}}}
\end{figure}

\begin{figure}
\includegraphics[scale=0.4]{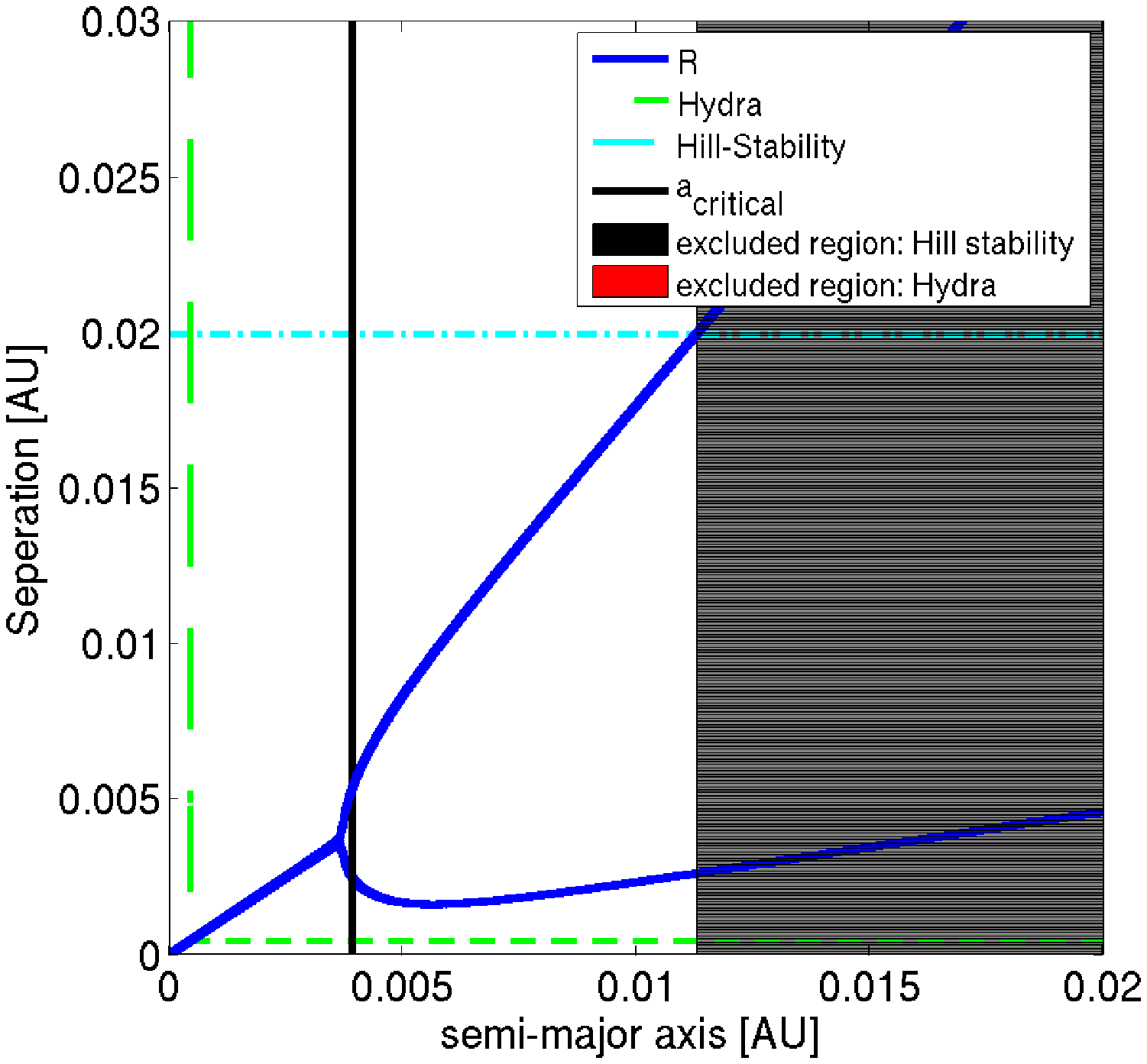}\includegraphics[scale=0.4]{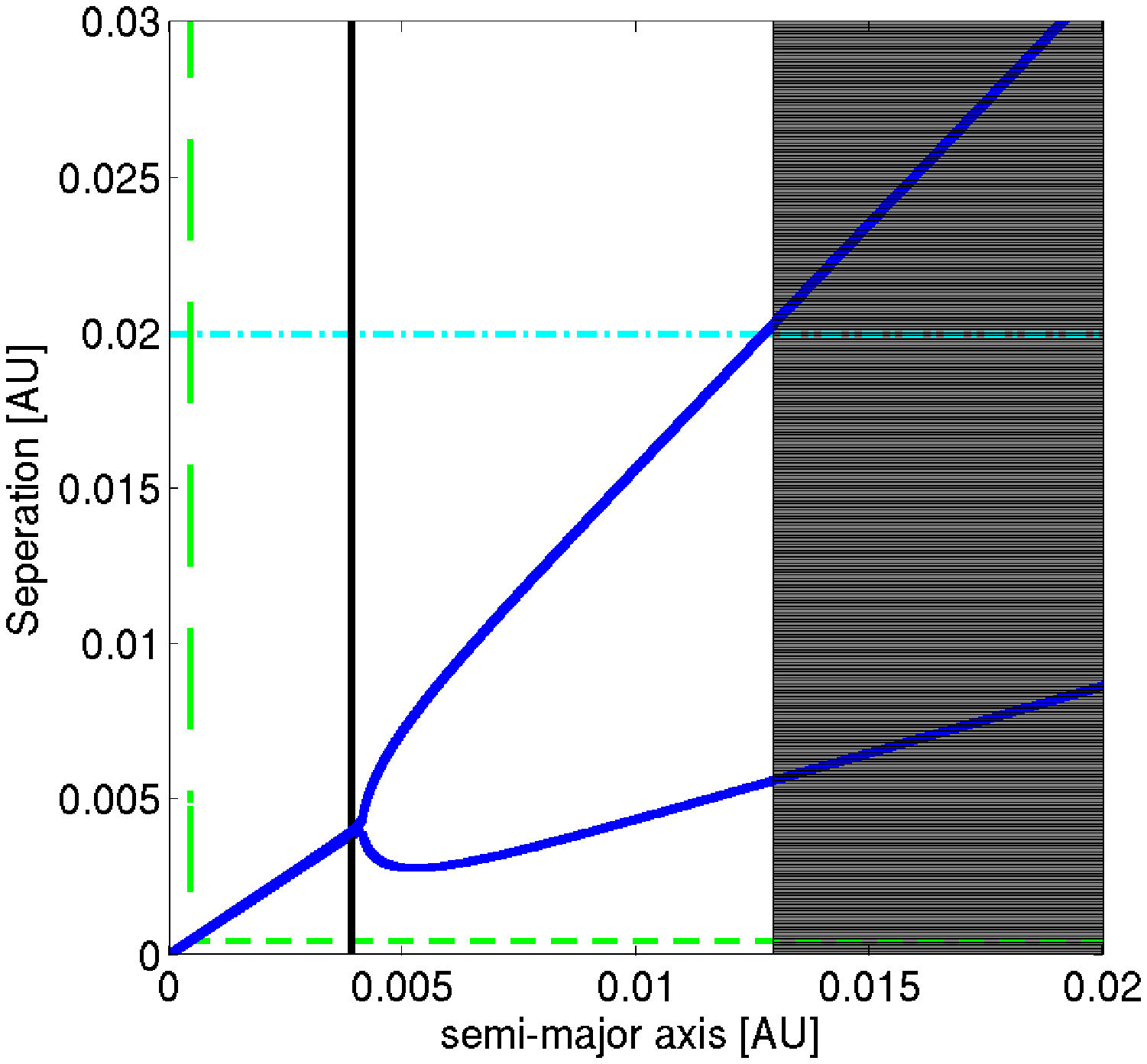}

\includegraphics[scale=0.4]{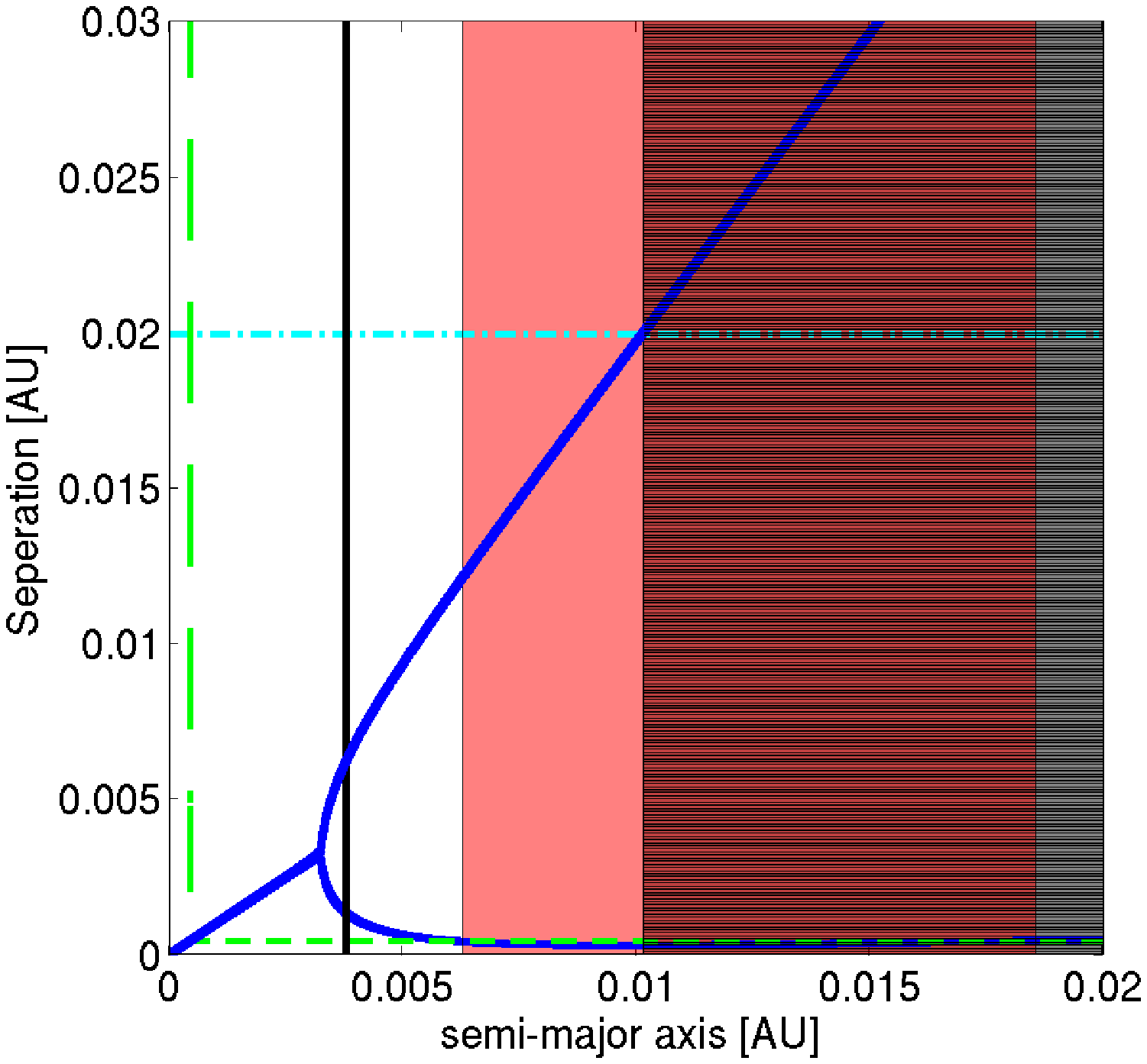}\includegraphics[scale=0.4]{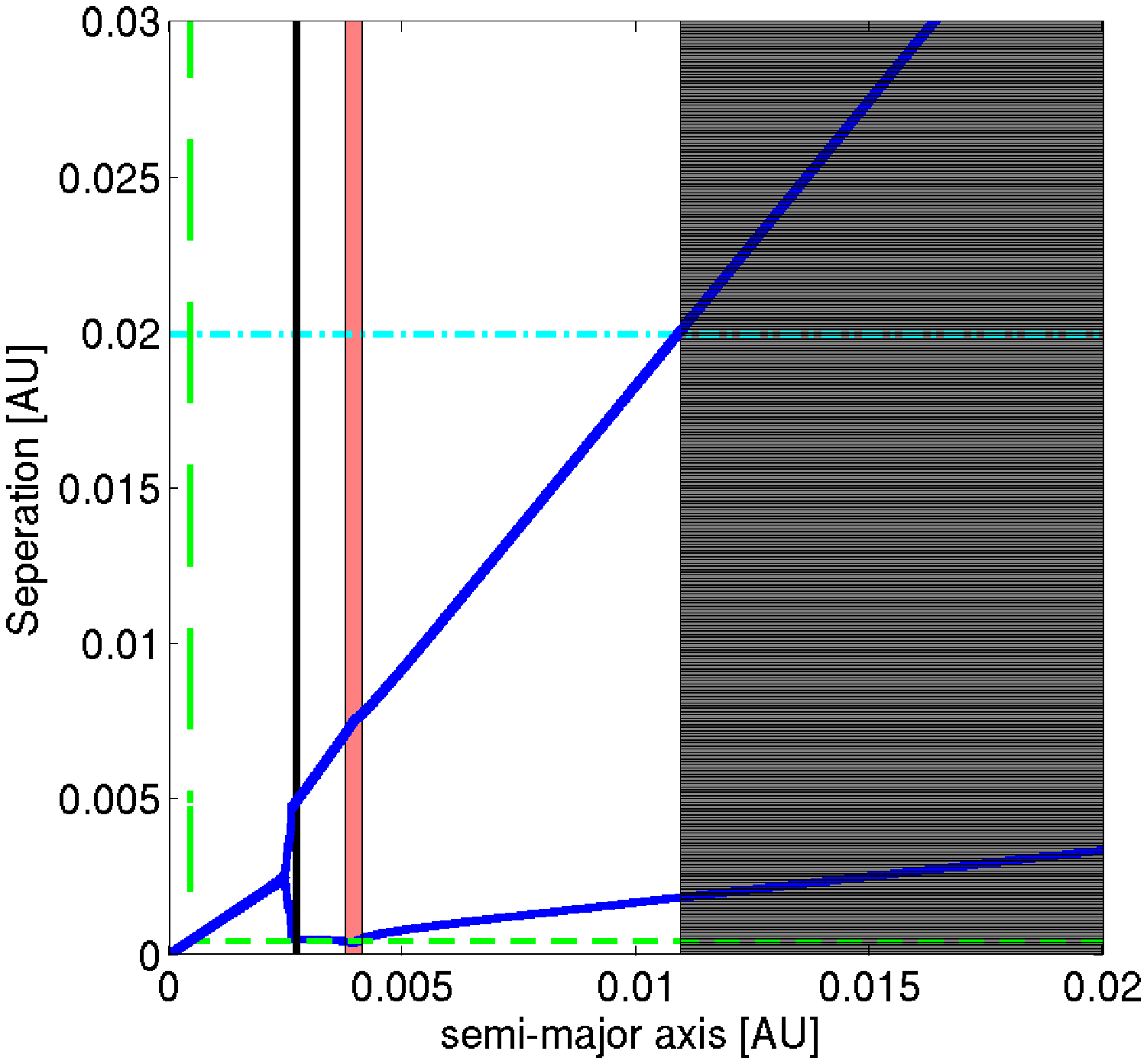}

\caption{{\footnotesize{\label{fig:regions}Allowed and excluded regions for
the existence of outer satellites and dust particles. The minimal
and maximal separations of satellites during their secular evolution
are shown as a function of their initial separation from the center
of mass of the Pluto-Charon system. Top left: Satellites on initially
co-planar circular orbits do not not secularly evolve up to distances
comparable with the critical separation. Beyond this point the eccentricities
and inclinations of such satellites are periodically highly excited.
The secular evolution drives the satellites through a range of separations,
enclosed by the blue solid lines. Satellites at sufficiently high
eccentricities become unstable as they cross the stability region
(Hill stability; $\sim0.5{\rm R_{H}}$ for retorgrade orbits; top
dashed vertical line) at apocenter, or if they cross the orbits of
the inner satellites at pericenter (bottom dashed line and left vertical
lines show Hydra SMA/separation). Regions in which such instabilities
occur are therefore excluded regions where outer satellites can not
survive. Top right: Similar but for satellites on initially inclined
orbits of $+10^{\circ}$. Bottom left: The same for satellites on
initially inclined (-20 degrees with respect to Pluto-Charon orbit;
$99^{\circ}$ with respect to the orbit of Pluto around the Sun) circular
orbits. Bottom right: The same for retrograde orbits (inital inclination
of $-55^{\circ}$).}}}
\end{figure}

In Fig. \ref{fig:timescales} we compare the satellite precession
timescales due to the inner-binary induced precession with the precession
timescales due to LK secular evolution. The opposite dependence of
these two precession timescales on the orbital separation give rise
to a critical separation at which they equalize, as derived in Eq.
\ref{eq:a_crit-1}. Beyond the critical separation,$a_{crit}$, at
which the timescales become comparable, satellites/rings become highly
susceptible to LK secular evolution. The effect of the inner binary
can be combined with the full secular equations of motion of the quadruple
system (Pluto-Charon+outer satellite+the Sun) to derive the full orbital
evolution of the outer satellite in time. We use such derivations
to find the maximal inclination and eccentricities attained by an
initially co-planar satellite orbiting Pluto-Charon on a circular
orbit, as a function of the satellite initial separation from Pluto
(Fig. \ref{fig:kozai-examples}). As can be seen, the LK precession
timescale is longer than the binary-induced precession timescale for
satellites at small separations, including the current locations of
Pluto satellites. LK-evolution induced eccentricity and inclination
oscillations are completely quenched in these regions below $a_{crit}$,
allowing for the stable orbits of the currently known satellites.
However, beyond this point, LK oscillations are not (or only partially)
quenched. Satellites in these regions experience large amplitude eccentricity
and inclination oscillations. In principle, satellites excited to
high eccentricities might even cross the orbits of the inner satellites
at peri-center approach or extend beyond the Hill stable region at
apocenter. The former case leads to strong interactions with the inner
satellites; such satellite will eventually destabilize the satellite
system or collide with one of the inner moons. In the latter case
the perturbations by the Sun will destabilize the orbit of the moon,
likely ejecting it from the system or sending into crossing orbits
with the inner moons (see examples in Fig. \ref{fig:kozai-examples}).
Regions in which such instabilities occur are therefore excluded regions
where outer satellites can not survive (Fig. \ref{fig:regions}). 

\begin{figure}
\includegraphics[scale=0.5]{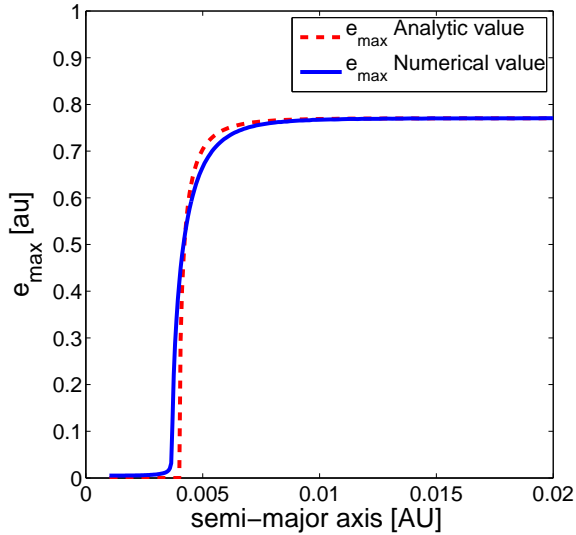}\caption{\label{fig:The-maximal-eccentricity}The maximal eccentricity attained
by an outer moon as a function of its semi-major axis (for satellites
on initially co-planar circular orbits). The maximal eccentricity
excited during the secular evolution depends on the the coupled effect
of the perturbations by the Sun (LK evolution) and the precession
induced by the Pluto-Charon inner binary. In the inner regions LK
evolution is quenched and the satellite keeps its initial eccentricity
and inclination, at the outer regions binary precession become negligible
and the maximal eccentricity is derived directly from the LK evolution.
In the intermediate regimes both process are important and the maximal
eccentricity can be derived numerically or be approximated analytically. }
\end{figure}

We can therefore exclude the existence of moons with some given orbital
parameters $a,\, i,\, e$ if they do not follow the stability criteria
\[
a_{hydra}<a[1-e_{max}(a,i,e)]\,\,{\rm {and}}\,a[1+e_{max}(a,i,e)]<R_{hill}
\]
where $e_{max}(a,i,e)$ is the maximal eccentricity of the moon during
an LK cycle (which can be analytically or numerically derived from
the coupled secular evolution equations of the orbital parameters;
including both binary precession and LK precession terms; see Fig.
\ref{fig:The-maximal-eccentricity}and appendix). Note that not far
beyond the critical separation the LK timescale becomes much shorter
than the binary-precession timescale, at which point the secular evolution
is completely dominated by LK evolution, and the binary precession
terms can be neglected, i.e. for such large separations one can use
the analytic solution for $e_{max}$ during LK oscillations, $e_{max}=[1-(5/3)\cos^{2}I]^{1/2}$
\citep{lid62} which is independent of the separation. This criterion
therefore provides a robust map of the excluded and allowed orbital
phase regions in which outer moons may exist in the Pluto-Charon system.
Moreover, the current stable configuration of the known inner satellite
system can therefore be used to constrain the region where farther
satellites can exist.

For initially circular co-planar moons with $a>a_{crit}$, $e_{max}\backsimeq0.78$
and we get $a_{crit}(1-e_{max})>a_{hydra}$, i.e. such moons never
evolve to hydra crossing orbits. This can be seen in Fig. \ref{fig:regions}
showing that initially regular moons on co-planar circular orbits
beyond $a_{m\_crit}$ evolve through LK cycles and can obtain high
inclinations and eccentricities, but their trajectories never cross
the orbits of the inner moons. Such moons, even if they were initially
formed on co-planar circular orbits (``regular'' moons) can not
sustain such orbits; they secularly evolve, and if observed they are
expected to be irregular moons, with non-negligible eccentricity and/or
inclination in respect to the Pluto-Charon orbit. Moreover, initially
regular outer moons can exist only in a limited outer region beyond
the critical separation, as satellites at even larger separations
cross the Hill stability radius. Large regions of the orbital phase
space are therefore excluded by these criteria for the existence of
moons. 

Our results therefore constrain the maximal extension where regular
moons and planetesimal disks can exist around the Pluto-Charon system.
We can therefore predict that any moons residing beyond the critical
separation, if such exist, will be eccentric/inclined irregular moons,
and we map the specific orbits allowed for such moons. These finding
also show that secular LK evolution can \emph{not} be neglected in
studies of the formation and evolution of the Pluto-Charon system
and its satellites, as hitherto (tacitly) assumed. 

\bibliographystyle{apj}

\appendix{}

\textbf{\Large{Appendix}}{\Large \par}

\section{Calculation of $a_{crit}$}

In the canonical two-body problem the bodies are treated as point
particles. Any deviation from the point mass approximation changes
the solution, and the orbital evolution. For example, if we change
the primary from a point mass to an oblate sphere the resulting orbit
of the secondary will be a precessing ellipse \citep{che+14}. The
same end result will be the case if we were to change the primary
point mass to a stable binary system. In this case the secondary orbits
the center of mass of the inner binary in a precessing Keplerian ellipse,
as is the case for Pluto-Charon and their coplanar zero eccentricity
moons \citep{lee+06}. In the following we describe the precession
rate for the more general case. Following similar derivations \citep{for+00,bla+02,nao+13,mic+14,ham+15}
we find that the precession rate of the tertiary in the quadrupole
expansion, the outer orbit, due to the inner orbit, is given by (the
inner Pluto-Charon orbit is constant): 
\[
\frac{dg_{m}}{dt}=3C_{2}\left\{ \frac{2\theta}{G_{PC}}\left[2+e_{PC}^{2}\left(3-5\cos2g_{PC}\right)\right]\right\} 
\]

\begin{equation}
+3C_{2}\left\{ \frac{1}{G_{2}}\left[4+6e_{PC}^{2}+\left(5\theta^{2}-3\right)\left(2+3e_{PC}^{2}-5e_{PC}^{2}\cos2g_{PC}\right)\right]\right\} 
\end{equation}
where 
\begin{equation}
C_{2}=\frac{Gm_{P}m_{C}m_{m}}{16\left(m_{P}+m_{C}\right)a_{m}\left(1-e_{m}^{2}\right)^{3/2}}\left(\frac{a_{PC}}{a_{m}}\right)^{2}
\end{equation}
the inner and outer angular momenta are 

\begin{equation}
G_{PC}=\frac{m_{P}m_{C}}{m_{P}+m_{C}}\sqrt{G\left(m_{P}+m_{C}\right)a_{PC}\left(1-e_{PC}^{2}\right)}
\end{equation}

\begin{equation}
G_{m}=\frac{m_{m}\left(m_{P}+m_{C}\right)}{m_{P}+m_{C}+m_{m}}\sqrt{G\left(m_{P}+m_{C}+m_{m}\right)a_{m}\left(1-e_{m}^{2}\right)}
\end{equation}

For Pluto-Charon case $e_{PC}=0$, due to tidal interaction, therefore
eq. \label{eq:binary_prec_general} reduces to 
\begin{equation}
\frac{dg_{m}}{dt}=6C_{2}\left(\frac{2\theta}{G_{PC}}+\frac{5\theta^{2}-1}{G_{m}}\right).\label{eq:binary_prec_reduced}
\end{equation}

Equation (\ref{eq:binary_prec_reduced}) sets a timescale for the
orbit precession. This timescale quenches Kozai cycle acting upon
the moon. In this case Kozai mechanism acting upon the following triple
system, the inner binary is the moon orbiting Pluto-Charon. In this
approximation Pluto-Charon considered to be a single object, due to
tidal interaction that sets their orbital parameters. The outer orbit
is the orbit that Pluto orbit the Sun. In order to find the SMA on
which the timescale are equal and hence find the critical SMA from
which the moon orbits is governed by Kozai mechanism we need to equate
the Kozai timescale and the precession timescale and solve for $a_{m}$,
the inner orbit SMA: 
\begin{equation}
6C_{2}\left(\frac{2\theta}{G_{PC}}+\frac{5\theta^{2}-1}{G_{m}}\right)=\left(\frac{dg_{m}}{dt}\right)_{Kozai}
\end{equation}
where $(dg_{m}/dt)_{Kozai}$ is given by 
\begin{equation}
\left(\frac{dg_{m}}{dt}\right)_{Kozai}\approx\frac{G^{1/2}m_{\odot}a_{m}^{3/2}}{\left(m_{P}+m_{C}+m_{m}\right)^{1/2}a_{p}^{3}\left(1-e_{P}^{2}\right)^{3/2}}
\end{equation}
where $e_{p}$ is Pluto's
orbital eccentricity and $G$ is Newton's constant. In the general
case we can solve this equation numerically. However, in the test
particle limit we can neglect the first term in the parenthesis of
the left hand side and solve for $a_{m}$ analytically. We can see
this fact if we compare the importance of the two term: 
\begin{equation}
\frac{2\theta/G_{PC}}{\left(5\theta^{2}-1\right)/G_{m}}=\kappa m_{m}\left(\frac{2\theta}{5\theta^{2}-1}\right)\label{eq:approx}
\end{equation}
where $\kappa$ is constant with units. We can see the in the test
particle limit $m_{m}\longmapsto0$ the first term in negligible except
an extreme case. In the case of a specific moon inclination that satisfies
the following condition 
\begin{equation}
5\theta^{2}-1=0
\end{equation}
which corresponds to 
\begin{equation}
\cos^{2}i=\frac{1}{5}.
\end{equation}

If the approximation is fulfilled then we can solve for $a_{m}$ and
get the critical $a_{m}$
\begin{equation}
a_{crit}^{5}=\frac{3}{8}\frac{a_{PC}^{2}a_{P}^{3}m_{P}m_{C}\left(1-e_{p}^{2}\right)^{3/2}\left(5\theta^{2}-1\right)}{\left(m_{P}+m_{C}\right)\left(1-e_{m}^{2}\right)^{2}m_{\odot}}.\label{eq:a_crit}
\end{equation}
Close inspection of (\ref{eq:a_crit}) we notice an interesting feature
of the dynamics. Due to the term $\left(1-e_{m}^{2}\right)$ in the
denominator moon orbits with high orbit eccentricity are more stable,
with respect to Kozai oscillations than circular ones, which is counterintuitive.

\section{Calculation of maximal eccentricity for initially co-planar circular
orbits}

The maximal eccentricity obtained by a moon during its long term evolution
can be derived by solving the full coupled secular evolution equations.
However, one can derive approximate analytic solutions in some specific
cases. In the following we derive the maximal eccentricity for a moon
on an initially co-planar circular orbit around the inner binary. 

In order to find the maximal eccentricity for initial nearly circular
orbit of the moon's orbit around Pluto-Charon we write the precession
equation of motion of the moon's orbit explicitly. The LK precession
rate in the quadrupole expansion in the nearly circular initial orbit
can be added to the binary precession by the inner Pluto-Charon system
to give 
\begin{equation}
\frac{dg_{m}}{dt}=\left(\frac{dg_{m}}{dt}\right)_{percession}+\left(\frac{dg_{m}}{dt}\right)_{Kozai}
\end{equation}

\begin{equation}
\frac{dg_{m}}{dt}=\frac{3}{8}\frac{G^{1/2}m_{P}m_{C}a_{PC}^{2}\left(5\theta^{2}-1\right)}{\left(m_{P}+m_{C}\right)^{3/2}a_{m}^{7/2}\left(1-e_{m}^{2}\right)^{2}}+\frac{3}{4}\frac{G^{1/2}m_{S}a_{m}^{3/2}}{\left(m_{P}+m_{C}\right)^{1/2}a_{p}^{3}\left(1-e_{P}^{2}\right)^{3/2}}\left(2-5\sin^{2}g_{m}\sin^{2}i\right)
\end{equation}
and from the definition of $a_{crit}$, see equation (\ref{eq:a_crit})
we can substitute and get

\begin{equation}
\frac{dg_{m}}{dt}=\frac{\left(3/2\right)\pi}{T_{Kozai}}\left[\left(2-5\sin^{2}g_{m}\sin^{2}i\right)+\frac{4}{3}\left(\frac{a_{crit}}{a_{m}}\right)^{5}\right]
\end{equation}
if we redefine $t$ into a more convenient coordinate 
\begin{equation}
\tau\equiv t\cdot\frac{\left(3/2\right)\pi}{T_{Kozai}}
\end{equation}
 
\begin{equation}
\frac{dg_{m}}{d\tau}=2-5\sin^{2}g_{m}\sin^{2}i+\frac{4}{3}\left(\frac{a_{crit}}{a_{m}}\right)^{5}=\left(2+\frac{4}{3}\left(\frac{a_{crit}}{a_{m}}\right)^{5}-5\sin^{2}g_{m}\sin^{2}i\right)\label{eq:percession}
\end{equation}
an immediate result is that as $a_{m}$ becomes larger than $a_{crit}$
the precession rate tends to the standard LK rate and therefore the
standard $e_{max}.$ The standard LK treatment for the test particle
problem leads to an additional constant of motion 
\begin{equation}
2=5\sin^{2}g_{m}\sin^{2}i
\end{equation}
(for derivation see, e.g. \citealp{val+06} ). In our system the correction
to the maximal eccentricity is due to the term that proportional to
$a_{m\_crit}/a_{m}$. Therefore the constant of motion is 
\begin{equation}
2+\frac{4}{3}q^{5}=5\sin^{2}g_{m}\sin^{2}i
\end{equation}
where we define $q\equiv a_{crit}/a_{m}$. From the former equation
we can get the relation between $g_{m}$ and $\sin i$. Together with
the secular evolution of the orbit eccentricity (for full equation
see \citealp{for+00,bla+02}) we get the following condition
\begin{equation}
\sin^{2}i>\frac{2+\frac{4}{3}q^{5}}{5}
\end{equation}
and by using the conservation of inner orbit angular momentum equation
\begin{equation}
\sqrt{1-e_{0}^{2}}\cos i_{0}=\sqrt{1-e_{max}^{2}}\sqrt{1-\frac{2+\frac{4}{3}q^{5}}{5}}
\end{equation}
\begin{equation}
\cos^{2}i_{0}=\left(1-e_{max}^{2}\right)\left(\frac{3-\frac{4}{3}q^{5}}{5}\right)
\end{equation}

\begin{equation}
1-\cos^{2}i_{0}\frac{5}{3-\frac{4}{3}q^{5}}
\end{equation}
and therefor the maximal eccentricity is 
\begin{equation}
e_{max}=\sqrt{1-\cos^{2}i_{0}\frac{5}{3-\frac{4}{3}q^{5}}}
\end{equation}
note that for 
\begin{equation}
3<\frac{4}{3}q^{5}
\end{equation}
the maximal eccentricity is ill defined; in these cases $a_{m}$ is
well below $a_{crit}$ and the secular LK evolution is completely
quenched.
\end{document}